\documentclass{article}

\usepackage[utf8]{inputenc}
\usepackage{authblk}
\usepackage{amsmath}
\usepackage{eufrak}
\usepackage{color,soul}
\usepackage{hyperref}
\usepackage{cite}
\usepackage{latexsym}
\usepackage{physics}
\usepackage{hyperref}
\usepackage{xcolor}
\usepackage{syntonly} 
\hypersetup{
    colorlinks=true,
    linkcolor=blue,
}

\title{Black Body Radiation in Moving Frames}

\author[ ]{Kamran Derakhshani\thanks{kderakhsh@physics.org}\\Zanjan, Iran}

\date{ }

\begin{document}
\maketitle

\begin{abstract}
    The problem of black body radiation, when measured by a moving observer, has a pivotal role in relativistic thermodynamics. Mutually, it depends on the thermodynamical definition of the thermal equilibrium and temperature of moving bodies, i.e. under a Lorentz transformation, and also in a gravitational field. Surprisingly, even after more than a century, relativistic thermodynamics is not a mature theory and is still an open problem without a consensus. This article is a brief review of the evolution of this theory with a special focus on the black body radiation in moving frames. As an application, we use the results in the most interesting topics of the quantum field theory in curved space: Hawking radiation, and Unruh effect. 
\end{abstract}
\textbf{Note}: We recast the adopted discussions of old-style papers in modern terminology, notations, and conventions as far as possible. In particular, we choose the signature $(-,+,+,+)$ for the metric and use the Einstein summation convention.
Throughout this paper, $\gamma = (1 - v^2/c^2)^{-1/2}$ is the Lorentz-Fitzgerald contraction factor, and the general-relativistic system of units ($c = 1$ and $G = 1$) is opted, unless explicitly stated. The spacetime indices are in Greek letters $\mu, \nu, ... = 0,1,2,3$,  and the space indices are in Latin letters $i, j, k, ... = 1,2,3$.

\section{Introduction}\label{intro}

 Although the subject of electromagnetic radiation of a moving body has been scrutinized since late nineteenth century, it was laid on its proper foundations after the discoveries of the distribution of black body radiation by Planck, and special relativity by Einstein. Naturally, these same giants and their ring of students were the first to explore the issue and hence the theory of relativistic thermodynamic was born. However, this was a premature birth and even today, after decades of proposing variety of models, it is not yet a fully-fledged theory and its very foundations are still under debate. As regards to the general relativistic thermodynamics (GRT) the situation is worse and the few attempts towards it are not as remarkable as that reserved by an established theory.
\\\\
This \textit{Report} is a review of the evolution of relativistic thermodynamics with a special focus on the problem of black body radiation from the point of view of an observer moving relative to it. It is a crucial subject of this theory, and is directly relevant to our future investigations.
In Section \ref{history} we report the history of the problem of thermodynamics of moving objects in a somewhat chronological manner. We begin with the earliest attention to this subject from the last quarter of the nineteenth century, within the framework of the study of electromagnetism of moving bodies. Then we explain the later periods, especially during the emergent of the theory of relativity, and expound the profound conflict between different points of view that, although not so hot today, is not settled   even yet.
Inasmuch as the main debate during decades is on the topic of special relativistic thermodynamics (SRT), and since the most viable version of SRT is that of Planck, Einstein, and their followers, we devote Section \ref{SRT} to an in-detail study of it. 
Section \ref{GRT} is allocated to the only version (as far as we are aware) of the theory of GRT that has almost solely been presented by Tolman. Since this theory approaches to the classical thermodynamics at the limit of flat spacetime, and also because it is one of the main pillars of our future works, we study it in more exquisite details with an emphasis on its most interesting applications to thermal equilibrium and black-body radiation.
In Section \ref{GRT_apps} we deal with some theoretical  applications of Tolman's GRT in finding local temperatures of the Hawking radiation and the Unruh effect. 
The \textit{Report} is wrapped up in the last Section (\ref{conclusions}) that summarizes the main points and highlights the most important results.

\section{Historical Overview}\label{history}
There are some traces of attention to the problem of thermodynamics of moving systems in the framework of electrodynamics of moving bodies during the later years of the nineteenth century in the works of Gibbs, Helmholtz, Boltzmann, Lorentz, etc. These endeavors extended to the earlier years of the twentieth century and culminates in the investigations of Hasen\"{o}hrl. But to stand upon correct footings it should pend for the theory of relativity to rise. From the very beginning of the advent of the theory of special relativity by Einstein in 1905, there was a rush to extend all fundamental theories of physics to the frame of inertial observers. Thermodynamics was not an exception to this ``relativitization'' movement. As expected, Planck, as the discoverer of the energy distribution of the black body radiation (which is also the starting point of the ideas of quantum and quantization in physics), apparently inspired by the work of one of his students (von Mosengeil), based the first serious milestone of the relativistic thermodynamics. Almost immediately, Einstein obtained the same results in some independent way. At least from the point of view of this \textit{Report}, their most important result is that the transformation law for the temperature of a moving body $(T)$ relative to that of the same body at rest $(T_0)$ is simply $T = T_0/\gamma$. In the next years, some other well-known physicists, such as von Laue, Pauli, and Tolman supported and enriched the theory. 
\\\\
However, due to some reasons, maybe the lack or impossibility of experimental evidence, the subject went to a dormant phase and up to the sixties it was almost forgotten. During that decade two events suddenly reawakened up the issue. First, a paper by Ott in 1963 seriously questioned the preassumptions of the Planck-Einstein's theory, and in an attempt to reformulate a new theory, he arrived at an irreconcilably different result, i.e. $T = \gamma T_0$. Although with some minor differences, Arzeli\`{e}s and a few others corroborated Ott's theory. More interestingly, years before them in 1952, in a few correspondences with von Laue, Einstein himself had thrown serious doubts on the Planck-Einstein's model. Second, the accidental discovery of the cosmic microwave background (CMB) in 1964 by Penzias and Wilson boosted the attempts to find at least an \textit{adhoc} formula for the temperature of the CMB (as a perfect black body radiation, as experimentally approved in the later years) while we along with Earth are moving within it. This led to the definition of an ``effective temperature'' that depended on the direction of observation and, anyway, was practically used by cosmologists to estimate the velocity of Earth relative to the CMB. The situation became more ridiculous when in later sixties Landsberg tried to show that, in contrast to mechanical variables, thermodynamic variables such as internal energy, temperature, entropy, etc. are not subject to Lorentz transformations and are indeed invariant, in particular  $T = T_0$. In the same years some people tried to clean up the mess by introducing a covariant formulation of thermodynamics. The pioneer of this road is van Kampen who in 1968 introduced an inverse-temperature four-vector. In his theory all thermodynamic variables may be treated as Lorentz invariants. Later on, this theory was clarified and developed by Israel in 1976. As far as we know, the last attempt on this track was made by Ford and O'Connell in 2013.
 
Clearly the main controversy deals with the \textit{transformation laws} for heat, temperature, and other thermodynamic quantities.
\\\\
According to what mentioned above, the history of relativistic thermodynamics inspired us to divide it into five distinguishable episodes, as follows.

\subsection{Pre-Relativistic era}
After the general acceptance of the Maxwell's theory of electromagnetism among physicists, it was not unexpected that some took attention to the problem of radiation by moving bodies. However, detailed investigations into their works, even up to earlier decades of the twentieth century, is a formidable task. Many of them was written in a non-English language (mostly in German) and no translation of them exists. As such, they are by today's standards presented in a tedious idiosyncratic style and sometimes include some typos or calculation errors. 
\\\\
Anyway, it seems that the first attempt to find the electromagnetic radiation of a moving body was done by Hasen\"{o}hrl, 1904 \cite{hasenohrl-1}. His work was based on some conclusions obtained by some previous physicists for the emission or absorption of radiation by flat surfaces in uniform motion. He applied their findings to the radiation in a moving cavity and showed that a part of the mechanical work done on a moving cavity is added to the radiation energy and this increase is, in first approximation, proportional to the squared velocity of the cavity as $(4/3)(v^2/c^2)E_0$, where $E_0$ is the radiation energy of the resting cavity. If we define mass by kinetic energy, this shows that the radiation energy has an ``apparent mass'' in addition to the ordinary mass. Then he calculated this mass to be $(8/3)(E_0/c^2)$. Although a little bit later the coefficient proved to be wrong  by Abraham \cite{abraham-1} and by Hasen\"{o}hrl himself \cite{hasenohrl-2} (the correct one is $4/3$), the idea that pure energy could have inertial mass was revolutionary at the time and finally culminated in the general and exact form of $E = mc^2$ by Einstein. What is important for us here is that an increase in the energy of radiation due to motion directly means an increase in its temperature, i.e. moving bodies seem to be hotter. This is the first explicit assertion of the dependence of temperature on speed. Hasen\"{o}hrl did not offer any calculation for this dependence and this task was left to be done by Planck.

\subsection{Planck, Einstein, etc.}

As mentioned above, the first one who treated the problem of thermodynamics of moving bodies in a special relativistic environment was von Mosengeil, 1907 \cite{mosengeil}. He himself used some previous studies on the radiation of moving mirrors done by a few people especially Abraham \cite{abraham-2}. His pioneering work was then slightly corrected, approved, developed, and also obtained in a totally different way by Planck \cite{planck}. Planck chose the principle of least action (in which he was interested so much) as the starting point, and using the principle of relativity reached the complete set of thermodynamic variables of a moving body. Immediately in chase of Planck, Einstein \cite{einstein-2} concluded the same results merely out of special relativity and the usual laws of thermodynamics. These seminal works, all published in 1907, consisted the ``standard'' theory of SRT and this theory was upheld and promoted by many, including von Laue (in his classical book ``Die Relativit\"{a}tstheorie", 1911 \cite{laue}), Pauli (in his famous article ``Relativit\"{a}tstheorie'', 1921 \cite{pauli}), and Tolman (in his book ``Relativity, Thermodynamics, and Cosmology", 1934 \cite{tolman}). Although there were some objections to this theory, and also it has never been experimentally tested it is still the most popular theory among the alternatives. 
\\\\
This is why we leave it for a while and allocate a complete section (Section \ref{SRT}) to explain it in this article.
\subsection{The Great Recession}
After the 1920s and due to some reasons, the Planck-Einstein's SRT was disregarded for more than forty years. After all it could not be directly tested. Even today we  can still not accelerate a macroscopic body to a remarkable fraction of the speed of light. Also, maybe it seemed to have no important application.
\\\\
In any case, in this period the subject of relativistic thermodynamics went idle. It was just been introduced in a few textbooks on relativity, eg. M\o ller, 1952 \cite{moller}.
\\\\
It is interesting that apparently the first one who challenged the Planck-Einstein's SRT was Einstein himself!(1952). During this period  in a few private letters to von Laue, that were never published\footnote{A sketch of these correspondences can be found in \cite{liu}}, Einstein designed a simple thought experiment (a relativistic Carnot cycle indeed) to prove that the correct transformations for temperature and transferred heat are
\begin{equation}
    \frac{T}{T_0} = \frac{Q}{Q_0} = \gamma
\end{equation}
The details of his reasoning was the same that would be independently used by Ott some years later, yet with much more discussion and development (cf. next section). In response, von Laue tried to show that any transfer of heat by a uniformly moving body is accompanied with a mechanical work to maintain its velocity constant, something on which Einstein never agreed. Einstein firmly believed that heat exchange did not require work. But this was not the end of the story. In 1953, in another letter to von Laue, Einstein stated that neither his opinion nor that of von Laue was ``optimum'' and the best-fit claim was that temperature was the notion of what was read by a comoving thermometer. By definition, this ``eigen temperature'' is Lorentz invariant. Since the entropy is also a Lorentz invariant \footnote{The justification of this statement is demonstrated in section \ref{PE}.}, the ``eigen heat'' is an invariant, too. Note that here the Lorentz invariance does not mean that it is measured the same by all inertial observers, but that temperature is what is measured in a comoving reference frame, and nothing else. This  necessitates implicitly that no reversible heat transfer is possible among bodies that are in relative motion to each other.
\\\\
This line of thought led Einstein to consider the \textit{eigen heat} as the four-vector 
\begin{equation}
    Q^{\mu} = Q\frac{dx^{\mu}}{ds}
\end{equation}
just like the energy-momentum vector. The time component of $Q^{\mu}$ is the pure heat (just like that of energy-momentum vector which is the proper mass), whereas its spatial components are the heat flows.
\\\\
All of these unpublished prophetic anticipations of Einstein were rediscovered and developed in the next decades; the subject of the next subsections. 
\subsection{Relativistic Thermodynamics Revisited}
The 1960s witnessed two important epochs that resurrected the debates on SRT. In the following, we briefly study them.

\subsubsection{Irreconcilable Alternatives to the Planck-Einstein's Theory}\label{alternatives}
\subsubsection*{a) Ott (1963)}
After decades, a posthumous paper by Ott in 1963 \cite{ott} started a new era of SRT. In his paper, Ott criticized the foundations of the Planck-Einstein's theory using a gedanken experiment. His approach is essentially the same as that of Einstein in 1952. Due to its importance, we state it here.
\\\\
Let we want to transfer an amount $dQ_0$ of heat from a reservoir at rest with temperature $T_0$ to a moving body with velocity $v$. We make this transfer by an intermediate body with proper mass $m_0$. So we connect this transporter to the reservoir in an isochoric (constant volume) process. According to special relativity, the mass of the transporter is increased by $dm_0 = dQ_0/c^2$. Now we accelerate it adiabatically to the velocity $v$ to make a thermal connection to the moving body. From the point of view of the rest frame the work that should be done on it is 
\begin{equation}
    dW_1 = (m_0 + \frac{dQ_0}{c^2})c^2(\gamma - 1).
\end{equation}
Now, the heat $dQ_0$ is transferred to the moving body. This heat transfer decreases the mass of the transporter by $(dm_0)c^2$. So to maintain its velocity at $v$ a ``driving force'' $F$ is needed. If its momentum is $P$, then
\begin{equation}
    \dot{P} =  \frac{d}{dt}(\gamma m_0v) 
\end{equation}
or at constant $v$
\begin{equation}\label{force}
    F = -\frac{dm_0}{dt}v\gamma.
\end{equation}
$F$ is a decelerating force, so the work done on the transporter  by it is
\begin{equation}
    dW_2 = Fvdt = -v^2\gamma dm_0 
\end{equation}
After the heat transfer, the transporter is again brought adiabatically to rest in contact with the reservoir. This time it accepts the work
\begin{equation}
    dW_3 = -m_0c^2(\gamma - 1).
\end{equation}
The overall process is a cyclic one. Thus the change in internal energy is zero, so $\delta Q = -\delta W$. In the rest frame of the reservoir the transferred heat is $dQ$, and according to the first law
\begin{align}\label{totalwork}
    dQ_0 - dQ &= -(dW_1 + dW_2 + dW_3)   \nonumber  \\
              &= -dQ_0(\gamma - 1) + dQ_0\gamma\frac{v^2}{c^2}. 
\end{align}
Therefore
\begin{equation}
    dQ = dQ_0/\gamma.
\end{equation}
So the Planck-Einstein's formula for the transformation of heat is again proved. However, Ott claims that the driving force $F$ has no physical reality. Of course, he does not deny that some braking force is needed to keep the velocity of the transformer constant. But he believes that it can not be in the form of (\ref{force}). By a few simple example, he shows that (\ref{force}) leads to ``strange consequences'' including negative kinetic energy and violation of the principle of energy conservation. Indeed, the transfer of the heat $dQ_0$ to the moving body exerts an impulse $d\pi$ on the transporter that must be accounted for in (\ref{force}). Using the formalism of four-dimensional quantities in the Minkowski spacetime, he obtains
\begin{equation}
    \dot{P} =  \frac{d}{dt}(\gamma m_0v) = F + \frac{d\pi}{dt}
\end{equation}
as the equation of motion, and 
\begin{equation}
   d\pi = \frac{dQ_0}{c^2}v\gamma.
\end{equation}
Note that when the transporter deliverers heat, $dQ_0 < 0$. This impulse does the work 
\begin{equation}
    dW_4 = vd\pi = \frac{dQ_0}{c^2}v^2\gamma
\end{equation}
that cancels out the work done by $F$, i.e. $dW_2$ in (\ref{totalwork}), then
\begin{align}
    dQ_0 - dQ &= -(dW_1 + dW_3)   \nonumber  \\
              &= -dQ_0(\gamma - 1).
\end{align}
So we have
\begin{equation}
    dQ = \gamma dQ_0
\end{equation}
which along with $S = S_0$ it results in
\begin{equation}\label{ott-temperature}
    T = \gamma T_0
\end{equation}
in sharp contrast to the Planck-Einstein's Theory\footnote{Historically, Eddington \cite{eddington} was the first one who suggested (\ref{ott-temperature}) in 1923. He argued that since $dS = dE/T$ and since entropy is a Lorentz invariant, so temperature transforms in the same way as energy, i.e. $T = \gamma T_0$. However, he never developed this more, because he believed that ``The transformation [of thermodynamic quantities] to moving axes introduces great complications without any evident advantages, and is of little interest except as an analytical exercise.''}.

\subsubsection*{b) Arzeli\`{e}s (1965)}
In spite of its significance, Ott's work on SRT was virtually ignored. However, two years later, and without any reference to Ott, Arzeli\`{e}s \cite{arzelies} deduced the same results, although from some different points of view and with some important differences on certain other thermodynamic variables, notably the internal energy. His paper ignited a hot and long debate on the issue and produced an abundant of interesting views and publications during 1960s and 1970s. Also, Gamba \cite{gamba}, while offering another simple approach to confirm the results obtained by Ott and Arzeli\`{e}s, scrutinized the source of ``error'' in the Planck-Einstein's theory and claimed that it could be traced in the relation for the transformation of volume, specifically that the correct formula was not $dV = dV_0/\gamma$ but 
\begin{equation}
    dV = \frac{dV_0}{\gamma(1-\beta cos\alpha)}
\end{equation}
where $\alpha$ is the angle between $v$ and the line of sight.

\subsubsection*{c) Landsberg (1966)}\label{Landsberg}
Of the hot debates on the transformation laws of thermodynamic variables, Landsberg's assertions are especially noticeable. In a one-and-a-half-page article in \textit{Nature} \cite{landsberg-1} he cast doubt on the Planck-Einstein's theory by the question ``Does a Moving body Appear Cool?''. He claimed that the transformation $T = T_0/\gamma$ has some unconvincing physical implications. Namely, temperature is a statistical concept and should not depend on the reference frame. Also, assume that two bodies are in equilibrium at rest at temperature $T_0$. Now if they are accelerated in isolation to different velocities , one would have $\gamma_1 T_1 = \gamma_2 T_2$ which is ``physically surprising''. These considerations inspired Landsberg to postulate that temperature is a Lorentz invariant, i.e for all inertial reference frames
\begin{equation}
    T = T_0.
\end{equation}
However, since the Planck-Einstein's theory is based on the standard form of the laws of thermodynamics, then this postulate induces back a gross modification of them. Specifically, he suggested that the first law for a moving body with uniform velocity $v$ and linear momentum $G$ is replaced by
\begin{equation}
    TdS = \gamma(dE + pdV - vdG).
\end{equation}
Also, the definition of temperature should be modified to
\begin{equation}
    \bigg(\frac{\partial S}{\partial E}\bigg)_{V,p} = \frac{\gamma}{T}.
\end{equation}
In this way, the Lorentz invariance of temperature is assured, and it directly gives
\begin{equation}
    dQ = dQ_0.
\end{equation}
Moreover, it is easily seen that the internal energy $U$ is also invariant. This generalization of thermodynamics was then elaborated and completed using statistical mechanics for a moving system in his next paper (1966)\cite{landsberg-2}. In response to some critiques and defences of the Planck-Einstein's theory, Landsberg (1967)\cite{landsberg-3} recalled another self-contradiction of the orthodox theory. Namely, if the moving body seems cooler, then the heat flow is from the rest frame to the moving frame. The same situation holds from the stance of the moving observer, i.e. in the reverse sense! Note that there is not such problem in the case of moving clocks, because that is completely symmetric and nothing is to be transacted between the frames. Landsberg asserted that this was the most direct argument in favor of a Lorentz-invariant temperature. 
However, in his same paper, he showed that to escape from the criticisms based upon the relativistic Doppler effect $\nu = \nu_0/\gamma$, the Planck distribution should be modified. Since in the rest frame of a black body the average occupation number of a mode of radiation  is $n_0 = f(h\nu_0/kT_0)$ and is an invariant, where $f$ is the Planck function, we must have
\begin{equation}
    n = n_0 = f(\gamma h\nu/kT).
\end{equation}
Moreover, to defend against some criticisms based on the equipartition theorem, Landsberg had to offer a new form of equipartition that is too elaborate to be mentioned here. Besides, he corrected the transformation of heat to be $Q = Q_0/\gamma$. This leads automatically to a new (generalized) form of the second law, i.e.
\begin{equation}
    \Delta S \ge \frac{\gamma \Delta Q}{T}.
\end{equation}
As can be seen, Landbergs' attempts to maintain the invariance of temperature leads to profound revisions in many of universally accepted laws and relations of thermodynamics. This is why Landsbergs' theory could not gain a general acceptance among physicists.
\\\\
Another important advance in this period is the suggestion of the inverse-temperature four-vector by van Kampen (1968) that due to its development mainly in the next decade, we delegate it to the next section.

\subsection{CMB-Motivated Studies, Anisotropic Temperature}\label{CMB}
What is really the relation between SRT and the cosmic microwave background? As a matter of fact, CMB is theoretically and experimentally confirmed to be a perfect black body radiation. Earth, along with the solar system, are moving within this sea of radiation. So we are dealing with a case of the black body radiation in a moving frame.
\\\\
Long ago, it was known that due to the relativistic aberration the intensity received by a moving observer flying within an otherwise isotropic radiation shows some anisotropy. Especially, the observer receives the maximum intensity in the parallel and the minimum in the antiparallel directions to her velocity. Assuming that the radiation is sort of a black body radiation, the same anisotropy is observed in the radiation temperature. 
\\\\
In 1968 Bracewell and Conklin \cite{bracewell} showed that an observer moving through a black body radiation with proper temperature $T_0$, measures again the spectrum of a black body, but with an anisotropic tempearture
\begin{equation}\label{CMB-temperature}
    T = \frac{T_0}{\gamma (1 - \beta cos\theta)}
\end{equation}
where $\beta = v/c$, and $\theta$ is the direction of measurement relative to the direction of motion, in the observer's rest frame. Also, Peebles and Wilkinson (1968) \cite{peebles}, using the fact that the number of photons that  a detector receives is a scalar, came to the same result, although with the preassumption that the distribution remains Planckian. Heer and Kohl (1968) \cite{heer} started from the fact, proved by Einstein in 1905, that $A/\nu$ of a plane wave is a Lorentz invariant, namely
\begin{equation}
    (A/\nu)^2 = (A_0/\nu_0)^2
\end{equation}
where $A$ and $\nu$ are the amplitude and the frequency, respectively. They also assumed in advance a Planck distribution and concluded that
\begin{equation}
    \nu/T = \nu_0/T_0
\end{equation}
and recalling the relativistic Doppler effect $\nu = \nu_0{\gamma (1 - \beta cos\theta)}^{-1}$ ended with the same conclusion as that of Bracewell and Conklin. Also, Henry \textit{et al.} \cite{henry} applied directly the Lorentz transformations to the Planck distribution and found that the distribution for a moving observer was still Planckian, but with an angle-dependent ``effective temperature'' with the same formula as in (\ref{CMB-temperature}). More importantly, they alerted that (\ref{CMB-temperature}) is by no means a transformation law for the temperature, i.e. $T$ is merely a parameter for the description of the Planck distribution in a moving frame. It is even possible to define more ``meaningful'' temperatures. For instance, if we plug this effective temperature in the Planck distribution, the exponential factor can be written as 
\begin{equation}\label{expfactor}
    exp\{\hbar\omega\gamma(1-\beta cos\theta)/kT_0\} = exp\{\gamma(\epsilon - \mathbf{p.v})/kT_0\}
\end{equation}
where $\epsilon$ is the photon energy, and $\bf{p}$ is its momentum. On the other hand, from statistical mechanics we know that if an observer slowly  moves in a fluid with velocity $\bf{v}$, then the Boltzmann factor will be
\begin{equation}\label{boltzmanfactor}
    exp\{(\epsilon - \mathbf{p.v})/kT_0\}.
\end{equation}
Now, taking (\ref{boltzmanfactor}) as the \textit{definition} of temperature, from (\ref{expfactor}) we have $T = T_0/\gamma$. Moreover, we can also rewrite (\ref{expfactor}) as 
\begin{equation}\label{fourexp}
    exp(p_\mu v^\mu/kT)
\end{equation}
where $p_\mu = (\epsilon/c,-\bf{p})$ designates the four-momentum of the photon, and $v^\mu$ = $\gamma(c,\bf{v})$ is the four-velocity of the observer. Choosing (\ref{fourexp}) as the definition of temperature gives $T = T_0$, namely temperature is a Lorentz invariant.
\\\\
The above discussion shows clearly that the deduction of a Planck distribution in a moving frame can by no means constraint the transformation law for the temperature. So, the only use of the effective temperature has been in the determination of Earth's velocity relative to the CMB.
\\\\

\subsection{Modern Studies}
After the 1960s, the subject of relativistic thermodynamics entered a moderate phase that is still continuing. In this period, most of studies have tried to prove or disprove the previous theories. Among them, only the four-vector formalism has something originally new  to offer, which we explain it in some details at the end of this section. Apart from that, the most important viewpoints in this period (in our opinion) are as follows.
\\\\
Aldrovandi and Gariel (1992)\cite{aldrovandi} claimed that the effective temperature (\ref{CMB-temperature}) is not only a mathematical shorthand and not merely a parameter, but a real transformation law. They regarded the observed dipole anisotropy of the CMB temperature as an experiment that could distinguish between the competing theories of the relativistic temperature and was obviously in favor of Ott's model. Briefly, if $\nu_0$ is the proper frequency of a sample of CMB photons received from direction $\theta_0$ in the CMB rest frame , then due to Doppler effect the Earth-borne detector senses them with frequency $\nu = \gamma(1 + \beta cos\theta_0)\nu_0$. As the energy of a photon is $h\nu$, so for the energy of this sample we have
\begin{equation}
    E = \gamma(1+\beta cos\theta_0)E_0.
\end{equation}
This relation can be used to define a thermodynamical temperature if the CMB can be considered as a ``confined'' system. According to the definition by Landsberg and Johns (1967)\cite{landsberg-4}, a confined system is a system of particles confined to a specified volume by some external forces. In equilibrium, such a system has a definite volume, pressure, and temperature. Aldrovandi and Gariel believe that this is the case for the CMB. They simply remind us of the relation $R(t)T = const.$  of the standard model of cosmology in a matter-dominated universe (where $R(t)$ is the length scale at time $t$) and that this is due to the interaction with gravitational field. So the CMB is in fact a confined system in a ``thermal bath'' of gravitation. Thus, noting that for an inertial reference frame, $\mathbf{v}$ and consequently $\theta$ are constants, and recalling that $S = S_0$ we have
\begin{equation}
    T = \bigg(\frac{\partial E}{\partial S}\bigg)_{V,\mathbf{v}} = \gamma(1+\beta cos\theta_0)\bigg(\frac{\partial E_0}{\partial S_0}\bigg)_{V,\mathbf{v}} = \gamma(1+\beta cos\theta_0)T_0
\end{equation}
which is the same as (\ref{CMB-temperature})
\footnote{Note that due to relativistic aberration $cos\theta = \frac{cos\theta_0 + \beta}{1+\beta cos\theta_0}$, so
\begin{equation}
 \gamma(1+\beta cos\theta_0) = \frac{1}{\gamma (1 - \beta cos\theta)}.    \nonumber  
\end{equation}}.
Hence they concluded that (\ref{CMB-temperature}) represents directly the Ott-Arzeli\`{e}s position. Then they showed that the Planck-Einstein's view led to a complicated relation for the same sample that was by no means similar to (\ref{CMB-temperature}) and was especially of no use in estimating the Earth's velocity relative to the CMB.
\\\\ 
Rather than arguments based on thermodynamics, Costa and Matsas (1995) \cite{costa} performed calculations in the context of relativistic quantum field theory. They used the Unruh-DeWitt detector as a thermometer and calculated its excitation probability and excitation rate, while moving through a thermal bath with temperature $T_0$, using standard quantum field theory methods. An Unruh-DeWitt detector is a two-level monopole that can be either in the ground state or in an excited state and can detect massless scalar particles (``spinless photons''). The obtained excitation rate (and following it, the particle number distribution $n(\omega)$) is dramatically non-Planckian, although its $v\to0$ limit is exactly that of a black body. Since this distribution can never be expressed in the black body form, so we have to generalize the concept of temperature. However, different prescriptions may result in opposite conclusions.
\\\\
The above-mentioned work of Costa and Matsas encouraged Landsberg and Matsas (1996)\cite{landsberg-5} to assert hopelessly that a universal and continuous Lorentz transformation of temperature does not exist, because there is no continuous function $T = T(T_0,v)$ that can convert the non-black-body distribution, detected by a moving Unruh-DeWitt thermometer, to a black-body one. In particular, all those famous relations presented by Planck-Einstein, Ott-Arzeli\`{e}s, and Landsberg are useless in this context. Also, operational definitions of temperature using different ``thermometers'' will lead to different functional dependencies. They believe that any manipulation of Lorentz transformations of thermodynamical variables also comes to doubtful results, unless the theory is made \textit{intrinsically} covariant (cf. next subsection). So the best we can do is to adopt the (\ref{CMB-temperature}) as just a ``directional'' temperature that can not be associated with a legitimate thermal bath (which is necessarily isotropic). They concluded that the proper (i.e. comoving) temperature alone is the only temperature of universal significance.

\subsubsection*{Covariant SRT}
Many people felt that the resolution to the riddle of relativistic thermodynamics might be in defining thermodynamic variables as four-dimensional beasts in a Minkowskian zoo. The first candidate would naturally be the temperature. But unfortunately temperature can not be considered as the time component of a four-vector, although the inverse temperature can.
\\\\
The first one who offered such an idea was van Dantzig (1939) \cite{dantzig}. He developed a formalism for the treatment of thermodynamics (and hydrodynamics) of matter in motion. The core of his theory is the introduction of a variable $\theta$, called ``thermasy'', that is the time integral of the absolute temperature and is defined by 
\begin{equation}
    d\theta = kTd\tau
\end{equation}
where $k$ is the Boltzmann constant. While $kT$ is not invariant, $d\theta$ is assumed to be. The invariance of $d\theta$ immediately results in the Planck-Einstein's theory. Using this notion, a \textit{temperature four-vector} can be defined as
\begin{equation}
    \Theta^\mu = \frac{dx^\mu}{d\theta} = \frac{1}{kT}u^\mu = \frac{1}{kT}(c,u^i)
\end{equation}
where $u^\mu$ is the four-velocity.
\\\\
However, the most promising approach to a covariant theory of thermodynamics is that of van Kampen (1968) \cite{kampen}. He initially showed that an extension of the definition of thermodynamic variables to moving systems, in such a way that the first and second laws of thermodynamics remain valid, led naturally, but not necessarily, to the Ott's theory while Planck-Einstein's thermodynamics implies an unsatisfactory formulation of the first law. Then he suggested a third formulation of relativistic thermodynamics that is covariant from the very beginning: Let $u_\mu = \gamma(1,\bf{u})$ is the four-velocity of the system, and $U_\mu = u_\mu U^0$ is its energy-momentum four-vector. Then we write the relativistic extension of the first law
\begin{equation}\label{covfirstlaw}
    dQ_\mu = dU_\mu + dW_\mu
\end{equation}
where $dW_\mu$ represents the mechanical energy and momentum transferred to another system. The four-vector $dQ_\mu$ is called the \textit{thermal energy-momentum transfer}. The component of this four-vector along the four-velocity in the rest frame
\begin{equation}
u^\mu dQ_\mu = dQ^0
\end{equation}
is called the proper \textit{heat supply}, which is an invariant. Therefore, to maintain the form of the second law
\begin{equation}
    dS = \frac{dQ^0}{T}
\end{equation}
the temperature $T$ must be defined as equal to the temperature $T^0$ in the rest frame. Thus in this covariant formulation of thermodynamics both $dQ$ and $T$ are scalars. The heat supply can be calculated using the first law (\ref{covfirstlaw}) as
\begin{equation}
    dQ^0 = u^\mu dU_\mu + u^\mu dW_\mu.
\end{equation}
Now, van Kampen applies this theory to a pair of comoving systems. In this case, the heat transfer is $dQ^0$ and the thermal momentum transfer is zero. Since $dQ_\mu$ is a four-vector, in any other inertial frame we have $dQ_\mu = u_\mu dQ^0$. In a similar manner, reversible transfer of the mechanical work $pdV^0$ transfers no momentum, so $dW_\mu = u_\mu dW^0$. Then the first law comes to
\begin{equation}
    u_\mu dQ^0 = dU_\mu + u_\mu dW^0.
\end{equation}
But we defined $U_\mu = u_\mu U^0$, so
\begin{equation}
    dU_\mu = U^0du_\mu + u_\mu dU^0.
\end{equation}
Then for constant velocity
\begin{equation}
    dQ^0 = dU^0 + dW^0.
\end{equation}
This is the  component  of (\ref{covfirstlaw}) parallel to the four-velocity, and simply the first law in the rest frame of the system. This shows that for comoving processes the covariant first law leads to the familiar classical one in the rest frame.
\\\\
Application of the van Kampen's theory to systems with different velocities is more delicate and interesting. Suppose a closed system of two interacting components $1$ and $2$. In this case $U^1_\mu + U^2_\mu$ is constant and $dW^1_\mu + dW^2_\mu = 0$. From the first law we then have $dQ^1_\mu + dQ^2_\mu = 0$. Obviously, for different velocities this does \textit{not} result in $dQ^1 + dQ^2 = 0$. That is, in general
\begin{equation}
  dQ^1 + dQ^2 \neq 0  
\end{equation}
and this means that \textit{``When thermal energy and momentum are transferred, the heat lost by one system is not necessarily equal in amount to the heat gained by the other system.''}. The reason is that the thermal energy-momentum four-vector $dQ_\mu$ is not decomposed into energy and momentum in the same way for different observers. They do not agree on the heat content of the transferred energy. 
\\\\
In the next step, van Kampen used the thought experiment of direct energy-momentum transfer between two black bodies with different velocities, and interestingly concluded for the variation of the entropy
\begin{equation}
    dS = \bigg(\frac{1}{T^0_1} - \frac{1}{T^0_2}\bigg)(\rho_2 - \rho_1) + (\gamma - 1)\bigg(\frac{\rho_1}{T^0_2} + \frac{\rho_2}{T^0_1}\bigg)
\end{equation}
where $\rho$'s are the energy densities of the black bodies and $v$ (implicit in $\gamma$) is the relative velocity of the two bodies. Note that due to the Stefan-Boltzmann law, $\rho$ increases with the proper temperature $T^0$. So the first term is always positive unless $T^0_1 = T^0_2$. Also the second term is always positive unless $\gamma = 1$. Thus \textit{``the entropy increases, unless both bodies have the same temperature \underline{and} velocity.''}. This means that any exchange of radiation between bodies with the same rest-frame temperatures, but different velocities, is an irreversible process. If we consider the maximality of the entropy ($dS = 0$) as the condition of equilibrium, we find that the equilibrium is reached when
\begin{equation}
    T^0_1 = T^0_2 \qquad and \qquad u^1_\mu = u^2_\mu .
\end{equation}
Therefore, bodies with relative velocities can not be in equilibrium even though their rest-frame temperatures are the same. 
\\\\
van Kampen himself showed also that the same thermodynamics could be deduced if we defined an inverse temperature four-vector $\beta_\mu = u_\mu/T$. Then, while keeping the first law in the covariant form (\ref{covfirstlaw}), we may then write the second law in the form
\begin{equation}
    dS = \beta^\mu dQ_\mu .
\end{equation}
Finally, van Kampen stated that if we accepted the covariant first law (\ref{covfirstlaw}) and distinguished between thermal energy-momentum transfer and heat, then the simplest choice would be to define the temperature as a scalar, that is $T = T^0$, and claimed that this choice was supported by statistical mechanics.
\\\\
van Kampen's covariant thermodynamics has relatively taken much attention so that a well-known relativist such as Werner Israel refined and developed it (1976) \cite{israel}. Nowadays it is called the van Kampen-Israel theory.
\\\\
The idea of the inverse temperature four-vector inspired some people to take their chance in relativistic thermodynamics. Among them, we choose the works of Nakamura . 
\\\\
Nakamura (2006) \cite{nakamura-1} believed that the van Kampen-Israel theory was one of the best solutions to the problem of relativistic thermodynamics. However, he claimed that it needed some slight corrections. According to Nakamura, van Kampen ignored the fact that the volumes of a finite object viewed by different inertial observers were different physical entities that were not related to each other by a Lorentz transformation, and this was a common mistake made by many physicists. In an attempt to correct this mistake, Nakamura achieved a clearer covariant definition of entropy
\begin{equation}
    dS = \beta V_0d\rho_0 - \beta pdV_0
\end{equation}
where $\beta = 1/kT$, $\rho_0$ is the energy density in the comoving frame, and $p$ is the pressure. In another paper in 2008 \cite{nakamura-2}, including a table of various theories up to date, Nakamura showed that the three main different views of relativistic thermodynamics (i.e. those of Planck-Einstein, Ott-Arzeli\`{e}s, and Landsberg) can be derived from the basic formulation of the van Kampen-Israel theory, depending on the way one decomposes the energy-momentum into the reversible and irreversible parts, and also on the definition of three-dimensional volume. 
\\\\
In his last paper on this subject (2009) \cite{nakamura-3}, Nakamura tried to make use of the concept of the inverse temperature four-vector to find the number density of photons of a moving black body radiation. On the base of standard statistical mechanical arguments, he concluded that the Boltzmann factor $exp(-\beta E)$ should be replaced with
\begin{equation}
    exp\big(-\beta_\mu P^\mu\big)
\end{equation}
where $P^\mu$ is the energy-momentum of the system. He finally found that
\begin{equation}
   \beta_\mu = \frac{u_\mu}{T^0} 
\end{equation}
where $u_\mu$ is the four-velocity of the black body. This is the same inverse temperature four-vector as in the van Kampen-Israel theory.
\\\\
We wrap up this historical overview with the work of Ford and O'Connell (2013) \cite{ford-1} which up to our knowledge is the last attempt explicitly made on SRT. They took the familiar expressions of quantum electrodynamics for the free electric and magnetic field operators in terms of the lowering and raising operators to find the correlation functions of the field fluctuations. Then, they applied the results to calculate the spectral distribution of black body radiation in rest frame as
\begin{equation}
    \rho(\omega,\mathbf{\hat k}) = \frac{\hbar}{(2\pi c)^3}\omega^3coth\frac{\hbar\omega}{2kT}
\end{equation}
where $\omega$ is the frequency of radiation and $\mathbf{\hat k}$ is its direction. This is the Planck distribution when the zero-point energy is included. To find the black body radiation in a moving frame, this time they began from the Lorentz transformations of the fields. Regarding the Lorentz transformation of the frequency (the Doppler shift) and that of the propagation vector (the aberration), they obtained
\begin{equation}
   \rho'(\omega',\mathbf{\hat k'}) = \frac{\hbar}{(2\pi c)^3}\omega'^3coth\bigg(\frac{\hbar\gamma(1+\mathbf{\hat k'}.\mathbf{v}/c)\omega'}{2kT}\bigg). 
\end{equation}
This expression shows that at $T = 0$ the spectral distribution is invariant. Furthermore, at finite temperatures this is the same form previously known in the context of the CMB (cf section \ref{CMB}). They believed that their derivation clarified the fact that $T$ is the invariant temperature in the rest frame of the black body, and therefore there has been no need to get into the ``confused and confusing'' question of how temperature transforms under Lorentz transformation. This conclusion remembers the opinion of Eddington in 1923.
\\\\
We end up here the historical overview of the SRT. However, since the most accepted theory seems to be that of Planck-Einstein, we expound it in more analytical details in the next section. Moreover, as it seems that the sole approach ever taken towards a general relativistic theory of thermodynamics is the Tolman's theory, we postpone it entirely to an allocated section (Section \ref{GRT}).

\section{Special Relativistic Thermodynamics}\label{SRT}
\subsection{von Mosengeil's Pioneering Work}
Historically, the first one who used special relativity in thermodynamics was one of the graduate students of Planck, named Kurd von Mosengeil. Unfortunately, before he published his doctoral dissertation, he died in a mountaineering accident. So there is no original information on his investigations. However, Planck and  Wien edited his draft for publication in the \textit{Annalen der Physik} in 1907. This posthumous paper \cite{mosengeil} was titled ``Theory of stationary radiation in a uniformly moving cavity" and contains the first relativistic expression for the temperature of a moving body as well as other equations of relativistic thermodynamics.

Here, to begin from the first point, and also as a tribute to von Mosengeil, we briefly cite his treatment.
\\\\
von Mosengeil starts off from the laws of reflection by moving mirrors previously derived by Abraham \cite{abraham-2}. Imagine a perfect mirror moving with velocity $\bf{v}$ relative to an isotropic source of light. If $J$ is the intensity of light (the rate of flow of electromagnetic energy in unit time passing through the unit area normal to the direction of radiation), then its \textit{specific intensity} (the radiation intensity per unit solid angle) is defined as $I = J/d\Omega$. von Mosengeil uses the Abraham's laws and some geometry to find the specific intensity in the reference frame of the mirror

\begin{equation}\label{intensity}
   I(\theta;v) = \frac{(1-\beta^2)^{8/3}}{(1-\beta cos\theta)^4}I_0
\end{equation}
where $v=\abs{\bf{v}}$, $\beta = v/c$, $I_0$ is the proper specific intensity of the source, and $\theta$ is the angle between $\bf{v}$ and the direction of the incident light beam, in the reference frame of the mirror.
\\\\
Now imagine a cavity with perfectly reflecting inside walls including some electromagnetic radiation in its local rest frame $K_0$, moving with velocity $\textbf{v}$ relative to an inertial reference frame $K$. From classical electrodynamics we know that if the intensity of the radiation  in $K_0$ is $I_0$, then its energy density $\epsilon_0$, its radiation pressure $p_0$, and its (center of mass) momentum density of radiation $g_0$ will be

\begin{equation}\label{EnergyPressureMomentum_1}
\begin{aligned}
    \epsilon_0 &= \frac{4\pi}{c}I_0  \\
    p_0 &= \frac{4\pi}{3c}I_0   \\
    g_0 &= 0  
\end{aligned}
\end{equation}

Using (\ref{intensity}) von Mosengeil finds the following relations for the same quantities for the moving cavity

\begin{equation}\label{EnergyPressureMomentum_2}
\begin{aligned}
    \epsilon &= \frac{4\pi}{c}I_0\frac{1+\beta^2/3}{(1-\beta^2)^{1/3}} \\
    p &= \frac{4\pi}{3c}I_0(1-\beta^2)^{2/3}    \\
    g &= \frac{16\pi}{3c^3}vI_0(1-\beta^2)^{-1/3}.      
\end{aligned}
\end{equation}

Now he takes a Carnot cycle very similar to that of Ott (cf. section \ref{alternatives}) without the constraint of constant volume. A body is used to transfer an amount of heat to a moving black body at uniform velocity $v$. In contact to a reservoir at temperature $T_0$ and during an isothermal process, it receives the heat $dQ_0$ and its volume increases by $dV$. If $dE_0 = \epsilon_0dV$ is the received energy by the transfer agent, and $dW_0 = -p_0dV$ is the work done on it, using (\ref{EnergyPressureMomentum_1}) we will have

\begin{equation}
    dQ_0 = \frac{16\pi}{3c}I_0dV.
\end{equation}

Now the transfer body is accelerated to the velocity $v$ in an adiabatic process and delivers the heat $dQ$ to the black body at temperature $T$. This time it also receives the mechanical energy $vgdV$ from the external force to maintain its velocity at $v$ during the heat transfer. So, using (\ref{EnergyPressureMomentum_2}), for the transferred heat we have

\begin{equation}
    dQ = \frac{16\pi}{3c}I_0(1-\beta^2)^{2/3}dV.
\end{equation}

In the last step, the transfer agent decelerates back to the rest in contact to the reservoir and completes the cycle. The entire process is reversible so $dS = 0$, where $S$ denotes the entropy. Then

\begin{equation}
    \frac{dQ}{T} = \frac{dQ_0}{T_0},
\end{equation}

or

\begin{equation}\label{temp}
    \frac{T}{T_0} = (1 - \beta^2)^{2/3}.
\end{equation}

By the Stefan-Boltzmann law $\epsilon_0 = aT_0^4$, where $a>0$ is a constant. Using (\ref{EnergyPressureMomentum_1}) and (\ref{temp}) in (\ref{intensity}) we conclude that

\begin{equation}
   I(\theta;v) = \frac{ac}{4\pi}\Big(\frac{T}{1-\beta cos\theta}\Big)^4.
\end{equation}

To find the specific intensity in the reference frame of the resting body we need to transform $v\longrightarrow-v$, $\theta\longrightarrow\theta'$, and $T\longrightarrow T'$. Then

\begin{equation}\label{I'}
   I' = \frac{ac}{4\pi}\Big(\frac{T'}{1+\beta cos\theta'}\Big)^4.
\end{equation}

We can use the well-known relation for the relativistic aberration

\begin{equation}\label{aberration}
    cos\theta' = \frac{cos\theta - \beta}{1 - \beta cos\theta}
\end{equation}

to write (\ref{I'}) in terms of $\theta$, i.e.

\begin{equation}\label{specific-intensity-1}
   I' = \frac{ac}{4\pi}T'^4\Big(\frac{1-\beta cos\theta}{1 - \beta^2}\Big)^4.
\end{equation}

On the other hand, we have the following transformation  for the intensities in the moving and resting frames  (derived by Einstein in his seminal paper on special relativity in 1905\cite{einstein-1})

\begin{equation}\label{total-intensity}
    J' = J\frac{(1 - \beta cos\theta)^2}{1 - \beta^2}.
\end{equation}

For the solid angle we know $d\Omega = sin\theta d\theta d\phi$, and since the direction of movement has been chosen along the z-axis, then $\phi' = \phi$. So (\ref{aberration}) is enough to show that

\begin{equation}
   d\Omega' = d\Omega\frac{1 - \beta^2}{(1 - \beta cos\theta)^2}.
\end{equation}

Recalling that $J = Id\Omega$, (\ref{total-intensity}) results in

\begin{equation}\label{specific-intensity-2}
    I' = I \frac{(1 - \beta cos\theta)^4}{(1 - \beta^2)^2}
\end{equation}

Since $\epsilon = \frac{4\pi}{c}I$, the Stefan-Boltzmann law reads

\begin{equation}
    I = \frac{ac}{4\pi}T^4.
\end{equation}

So (\ref{specific-intensity-2}) gives

\begin{equation}\label{specific-intensity-3}
I' = \frac{ac}{4\pi}T^4\frac{(1 - \beta cos\theta)^4}{(1 - \beta^2)^2}    
\end{equation}

A comparison of (\ref{specific-intensity-1}) to (\ref{specific-intensity-3}) takes us to the final result

\begin{equation}
    T' = T/\gamma.
\end{equation}

From the point of view of this \textit{Report}, this is the most important result of the von Mosengeil's work. It simply means that a black body in uniform motion seems cooler.
\\\\

\subsection{Planck-Einstein's Theory of Special Relativistic Thermodynamics }\label{PE}
In this \textit{Report} we study the Planck-Einstein's SRT more emphatically for  a few reasons. It is the most agreed one in the general public of physicists, although its acceptance may be in part due to its giant innovators and supporters such as Planck, Einstein, Pauli, von Laue, Tolman, etc., and the lack of any highly-celebrated scientist among its opponents. Moreover, one should not forget that this theory has been used in such a serious experimental context as the cosmological microwave background radiation.  
\\\\
All over his life, Planck used to praise the crucial role of the principle of least action in all parts of physics. He believed in it as the King of Principles in physics from which mechanics, electrodynamics, and thermodynamics could be deduced. However, he was also aware of its limitations. He noticed that the principle of least action leads to those theories because for ordinary (ponderable) bodies we can decompose the total energy into kinetic and internal energies. This decomposition is not generally possible. For example for a body containing radiant energy it can not be allowed. In such cases, we have to resort to the principle of relativity. In the following, we try to explain his line of reasoning and glean the parts directly relevant to the topic of this \textit{Report}.
\\\\
For any approach to the relativistic thermodynamics we have to know the entropy of a moving body. Fortunately, there is a unanimous view on this issue among physicists including Planck. Keeping in mind that the entropy is just another name for probability, and probability can not depend on the choice of the reference frame, we conclude that entropy is a Lorentz invariant, i.e. $S' = S$. Nevertheless, Planck prefers to prove it directly, without the introduction of the concept of probability, and indeed uses a \textit{reductio ad absurdum}, as follows.
\\\\
Consider a body at rest in the reference frame $K$ is brought to rest in the reference frame $K'$, which is moving with a uniform speed $v$ relative to $K$, through a reversible adiabatic process. So, for the initial and final entropies in $K$ and $K'$ we have $S_i = S_f$ and $S'_i = S'_f$. Now if $S'_i \neq S_i$, say $S'_i > S_i$, this means that for the moving observer the entropy of the body is greater. As there is no preferred inertial reference frame, the same must be true from the viewpoint of the moving observer. So, at the end of the process we must have $S_f > S'_f$. This obviously leads to a contradiction. In the same way, $S'_i < S_i$ also comes to a contradiction. Therefore, there is no possibility other than $S'_i = S_i$, or in general $S' = S$. This means that entropy is a Lorentz invariant.
\\\\
Then Planck considers the case of moving black body and uses the relations (\ref{EnergyPressureMomentum_2}), derived by von Mosengeil, to find another relations for radiation pressure $p$ and translational momentum $G$ in terms of energy density $\epsilon$

\begin{equation}\label{pG}
\begin{aligned}
    p &= \frac{\epsilon}{\gamma^2(3+\beta^2)}   \\
    G &= \frac{4\beta}{c(3+\beta^2)}V\epsilon    
\end{aligned}
\end{equation}

where $V$ is the volume of the cavity. 
\\\\
All changes of the system is assumed to be reversible, so the state of the body is completely determined by the variables $v, V$ and the temperature $T$.
By the first law of thermodynamics $dE = dQ + dW$ where $E = \epsilon V$ is the energy of the radiation, $dQ$ is the heat delivered to, and $dW$ is the external mechanical work done on it. In this case, $dW$ consists of the translational work $vdG$ and the work $-pdV$ done by the pressure . 
\\\\
By the second law of thermodynamics for reversible processes $dS = dQ/T$, where $S$ is the entropy of the radiation. So

\begin{equation}
    dE = TdS - pdV + vdG
\end{equation}

Using (\ref{pG}) we have

\begin{equation}
    dS = \frac{1}{T}\Big\{d(\epsilon V) + \frac{\epsilon}{\gamma^2(3+\beta^2)}dV -vd\Big(\frac{4\beta}{c(3+\beta^2)}V\epsilon\Big)\Big\}.
\end{equation}

However, $dS$ must be a complete differential of the state variables $v, V$, and $T$. Therefore

\begin{equation}
    \epsilon = \frac{1}{3}a(3 + \beta^2)\gamma^6T^4
\end{equation}
and
\begin{equation}\label{S}
    S = \frac{4}{3}aV\gamma^4T^3
\end{equation}

where $a$ is a constant. Note that in the local rest frame the Stefan-Boltzmann law for the black body radiation holds, i.e. we must have $\lim_{v\to0}\epsilon = aT^4$. So the constant $a$ is the same as the Stefan-Boltzmann constant. Using the obtained value for $\epsilon$, we find (by the definition of $\epsilon$ and (\ref{pG}))

\begin{equation}\label{EpG}
\begin{aligned}
    E &= \frac{1}{3}a(3 + \beta^2) \gamma^6T^4V     \\
    p &= \frac{1}{3}a\gamma^4T^4    \\
    G &= \frac{4a}{3c^2}v\gamma^6T^4V.     
\end{aligned}
\end{equation}

Using $\epsilon_0 = \frac{1}3p_0$ in the first law $dE_0 = T_0dS_0 - p_0dV_0$ for the black body at rest we will have $S_0 = \frac{4}{3}aT_0^3V_0$. Now, since $S = S_0$  and $V = V_0/\gamma$, then (\ref{S}) results

\begin{equation}
    T = T_0/\gamma.
\end{equation}

We observe that Planck also obtains the same results as those of von Mosengeil, yet in a more direct and easier way.
\\\\
A little bit later in the same year 1907, in a paper on the consequences of the principle of relativity, and in a section on the mechanics and thermodynamics of systems, Einstein \cite{einstein-2}, having chosen the velocity $v$, the volume $V$, and the energy $E$ of a moving body as the state variables, and using only the principle of relativity, obtained the following formulas for the energy and momentum of a body with mass $m$, energy $E_0$ in the local rest frame, and moving with uniform velocity $v$

\begin{align}
      E &= \gamma(mc^2 + E_0)  \nonumber  \\
      G &= \gamma(mc^2 + E_0)\frac{v}{c^2}.    \nonumber
\end{align}
 
Now, if in the local rest frame an external mechanical force exerts a pressure $p_0$ on the body with volume $V_0$, then

\begin{equation}
\begin{aligned}
      E &= \gamma(mc^2 + E_0) +  \gamma\frac{v^2}{c^2}p_0V_0   \\
      G &= \gamma(mc^2 + E_0 + p_0V_0)\frac{v}{c^2}.   
\end{aligned}
\end{equation}

Then, directly from the Lorentz transformations for the coordinates, he obtained  the transformation law for the volume of the moving body, and since according to the same transformations force and area transform alike (they are divided by $\gamma$), he concludes that the pressure is a Lorentz invariant, i.e.

\begin{equation}
\begin{aligned}
V &= V_0/\gamma   \\
p &= p_0
\end{aligned}
\end{equation}

Now we apply the above equations in the first law of thermodynamics for a moving body, i.e.
\begin{equation}
dE = dQ - pdV + vdG   \nonumber
\end{equation}
to obtain
\begin{equation}
dQ = (dE_0 + p_0dV_0)/\gamma  \nonumber
\end{equation}
or
\begin{equation}
dQ = dQ_0/\gamma 
\end{equation}
\\\\
For the entropy of the moving body he quoted verbatim the line of reasoning of Planck and deduced $S = S_0$. Since according to the second law of thermodynamics  $dQ = TdS$ for reversible processes, then the temperature follows the same transformation law as that of heat. So  
\begin{equation}
T = T_0/\gamma  
\end{equation}
in exact accordance to the results obtained by von Mosengeil and Planck.

\section{General Relativistic Thermodynamics}\label{GRT}
As far as we are aware, the first attempt to invent a complete thermodynamical theory in a curved spacetime has been made by R. C. Tolman. In a series of papers published since 1928 up to 1930 \cite{tolman-1,tolman-2,tolman-3,tolman-4}, Tolman postulated a general-relativitistically modified form of the two laws of thermodynamics.

His proposed first law is nothing but the law of conservation of energy-momentum in its covariant form and was derived using the fundamental equation of relativistic dynamics.

The second law of Tolman's thermodynamics is however a new innovation based on the definition of an entropy four-vector. He showed that the total entropy of a finite closed system never decreases. 

Then he used his theory on the Einstein's closed static universe (that was then still a tenable cosmological model), and more importantly, to the black body radiation. Also, he deduced a new concept of thermal equilibrium and the interesting result that in a gravitational field the proper temperature of a body in thermal equilibrium depended on its position.

Apparently there is no other specific and independent investigations on the issue after Tolman, and quite contrary to the case of SRT, Tolman's GRT has become the standard theory of GRT in the community of physicists, although it has not used in so many applications and also there has been no experimental evidence to prove or disprove it, as yet.

In this section, first we concisely explain the concept and formalism of Tolman's theory in modern notations and then present its application to the problem of black body radiation.
\subsection{Tolman's Theory of General Relativistic Thermodynamics}
Motivated by a work  of Lenz on the equilibrium between radiation and matter in Einstein's closed universe, Tolman suggested the covariant form of thermodynamic laws as the following.
\subsubsection{The First Law}
As the first law of the GRT, Tolman takes the law of conservation of energy in general relativity 
\begin{equation}
    T^{\mu \nu}_{\quad;\nu} = 0
\end{equation}
where $T^{\mu \nu}$  is the energy-momentum tensor. Using the tensor density ${\mathfrak {T}}^{\mu}_{\;\nu}=T^{\mu}_{\;\nu}\sqrt {-g} $  we have

\begin{equation}
    {\mathfrak {T}}^{\mu}_{\;\nu;\mu} = 0
\end{equation}
or\cite{eddington}
\begin{equation}
    {\mathfrak {T}}^{\mu}_{\;\nu,\mu} - \frac{1}{2} {\mathfrak {T}}^{\alpha\beta}g_{\alpha\beta,\nu} = 0
\end{equation}
and by integrating over a finite four-volume $V$
\begin{equation}\label{first_law}
    \int_{V}({\mathfrak {T}}^{\mu}_{\;\nu,\mu} - \frac{1}{2} {\mathfrak {T}}^{\alpha\beta}g_{\alpha\beta,\nu})dV = 0
\end{equation}
This equation is the first law of Tolman's GRT.

\subsubsection{The Second Law}
The second law of Tolman's GRT is not as easy and straight forward. It will be a postulate satisfying two requirements:
\\\\
1) It must be true in all coordinate systems.\\
2) It must approach the second law of classical thermodynamics at the limit of flat space-time.\\

Towards this goal, first we define an \textit{entropy four-vector} at any point of space-time as the following\footnote{Its other names are \textit{entropy flux} and \textit{entropy current} four vector.}
\begin{equation}\label{entropy_vector}
    S^{\mu}=\phi_0 \frac{dx^{\mu}}{d\tau}
\end{equation}
where $\tau$ is the proper time, $\phi_0$ is the \textit{proper density of entropy} at the point, and $dx^{\mu}/d\tau$ is the \textit{macroscopic four-velocity} of matter at that point, say $(1,u,v,w)$. 
\\\\
Now we postulate that over any four-dimensional volume $V$ that includes an isolated body we have

\begin{equation}\label{second_law}
    \int_VS^{\mu}_{\;;\mu}\sqrt{-g} dV \geq 0
\end{equation}

as the second law of Tolman's GRT.

\subsubsection{Testing Against the Requirements}
Do these results fulfill the above-mentioned requirements?
First, we note that the integral in the first law (\ref{first_law}) is not a tensor, however it holds for all sets of coordinates. As to the second law (\ref{second_law}), $S^{\mu}_{\;;\mu}$ and $\sqrt{-g}dV$ are both scalars so the equation itself is a tensor equation of rank zero. Thus, both laws of GTR meet the principle of covariance.
Second, to show that these laws reduce to the laws of classical thermodynamics we consider a system of Galilean coordinates $x,y,z,t$. Then $\sqrt{-g} = 1$ and $S^{\mu}_{\;;\mu} = S^{\mu}_{\;,\mu}$. So by the definition of the entropy vector (\ref{entropy_vector})

\begin{equation}
   {S}^{\mu} = \phi_0(\frac{dt}{d\tau}, \frac{dx}{d\tau}, \frac{dy}{d\tau}, \frac{dz}{d\tau}) = \phi_0(\frac{dt}{d\tau}, \frac{dx}{dt}\frac{dt}{d\tau}, \frac{dy}{dt}\frac{dt}{d\tau}, \frac{dz}{dt}\frac{dt}{d\tau}).
\end{equation}

Noting that $dt/d\tau = \gamma$ we have

\begin{equation}
   {S}^{\mu} = \gamma\phi_0(1,u,v,w) 
\end{equation}

However, the statistical interpretation of entropy in terms of probability (and also the Planck's reasoning mentioned above in section \ref{PE} shows that the entropy of an infinitesimal region is a Lorentz invariant. However, since the volume decreases by a factor of $\gamma$ the entropy density is multiplied by $\gamma$, under a Lorentz transformation, i.e. $\phi = \gamma \phi_0$. Then

\begin{equation}
S^{\mu}_{\;,\mu} = \frac{\partial \phi}{\partial t} + \frac{\partial}{\partial x}(\phi u) + \frac{\partial}{\partial y}(\phi v) + \frac{\partial}{\partial z}(\phi w)
\end{equation}

Thus, for an isolated system (no convection of material or flow of heat) and in flat space-time the second postulate (\ref{second_law}) will reduce to

\begin{equation}
   \int_V\frac{\partial \phi}{\partial t}dxdydzdt = \int^{t'}_t\frac{dS}{dt} dt = (S' - S) \geq 0
\end{equation}

where $S$ and $S'$ are the total entropies of the system at times $t$ and $t'$, respectively.
Also, in flat space-time the first law of GRT (\ref{first_law}) for $\mu = 0$ will reduce to

\begin{equation}
   \int^{t'}_t\frac{dU}{dt} dt = (U' - U) = 0
\end{equation}

where U is the total energy of the isolated system.
\\\\
Hence, the postulates of Tolman's GTR fulfill also the second of the two necessary requirements, namely, reducing to ordinary thermodynamics in flat space-time.

\subsection{Some Applications of Tolman's GRT}

In sequel, Tolman tries to apply his theory of GRT to some important cases. His main focus is the Einstein's closed universe that was still debatable as a cosmological model at that time. Although his work on this problem are pedagogically and historically important, we will expound here some other applications that are more directly relevant to the topic of this \textit{Report}. So after a brief preparatory introduction, we focus on the subject of thermal equilibrium and the special case of the black-body radiation.

\subsubsection{Preamble}

To discuss thermodynamics in the realm of general relativity, we need to somehow relate the thermodynamic quantities to the metric. We mean specifically some explicit relations between the proper energy density $\rho_0$ and the proper pressure $p_0$ from one hand, and the metric tensor $g_{\mu\nu}$ on the other hand.\\

Consider a \textit{static} distribution of a \textit{perfect fluid} having \textit{spherical symmetry}. The most general form of a metric that is static and has spherical symmetry in polar coordinates is 

\begin{equation}\label{line_element}
    ds^2 = -e^{\mu(r)} dt^2 + e^{\nu(r)}dr^2 + r^2d\theta^2 + r^2sin^2\theta d\phi^2 
\end{equation}
where we choose a general coordinate system $(x^0,x^1,x^2,x^3)$ with $x^0$  as the time coordinate. Remember that the metric signature is $(-,+,+,+)$, and the general-relativistic system of units ($c = 1$ and $G = 1$) is chosen, so $ds^2 = -d\tau^2$. Due to staticity and spherical symmetry $\mu$ and $\nu$ are functions of $r$ only, where $r$ is the distance to the origin.\\

So the nonzero components of the metric, their reverses, and the metric determinant respectively are

\begin{equation}\label{metric}
\begin{aligned}
g_{00} &= -e^\mu, & g_{11} &= e^\nu, & g_{22} &= r^2, & g_{33} &= r^2sin^2\theta,  \\
g^{00} &= -e^{-\mu}, & g^{11} &= e^{-\nu}, & g^{22} &= \frac{1}{r^2}, & g^{33} &= \frac{1}{r^2sin^2\theta}, \\
\sqrt{-g} &= e^{\frac{\mu+\nu}{2}}r^2sin\theta . 
\end{aligned}
\end{equation}

Now we calculate the nonzero components of the Ricci curvature tensor, and the Ricci scalar

\begin{equation}
\begin{aligned}
R_{00} &= e^{\mu-\nu}\Big (\frac{\mu''}{2} + \frac{\mu'^2}{4} -  \frac{\mu'\nu'}{4} + \frac{\mu'}{r} \Big )   \\
R_{11} &= -\frac{\mu''}{2}  - \frac{\mu'^2}{4} + \frac{\mu'\nu'}{4}  + \frac{\nu'}{r}   \\
R_{22} &= e^{-\nu}\Big[\frac{r}{2}(\nu' - \mu') -1\Big] + 1  \\
R_{33} &= sin^2\theta R_{22}          \\
R &= -e^{-\nu}\Big[\mu'' + \frac{\mu'^2}{2} - \frac{\mu'\nu'}{2} + \frac{2}{r}(\mu'-\nu')  + \frac{2}{r^2}(1 - e^\nu) \Big]  
\end{aligned}
\end{equation}

while $'=d/dr$. Using the Einstein's field equation

\begin{equation}
R_{\mu\nu} - \frac{1}{2}Rg_{\mu\nu} = 8\pi T_{\mu\nu}
\end{equation}

we can obtain the nonzero components of the energy-momentum tensor as

\begin{equation}\label{emtensor-1}
\begin{aligned}
8\pi T_{00} &= \frac{e^{\mu}}{r^2}\Big \{e^{-\nu}(r\nu' - 1) + 1 \Big \}     \\
8\pi T_{11} &= \frac{\mu'}{r} + \frac{1}{r^2}(1 - e^\nu)    \\
8\pi T_{22} &= r^2e^{-\nu}\Big\{\frac{(\mu'-\nu')}{2r} + \frac{1}{2}(\mu'' + \frac{\mu'^2}{2} - \frac{\mu'\nu'}{2})\Big\}\\
8\pi T_{33} &= 8\pi T_{22}sin^2\theta     
\end{aligned}
\end{equation}
\\\\
On the other hand, for a perfect fluid we have

\begin{equation}
T_{\mu\nu} = (\rho_0 + p_0)\frac{dx_\mu}{d\tau}\frac{dx_{\nu}}{d\tau} + p_0g_{\mu\nu}
\end{equation}

Remember that for a static system the macroscopic velocities vanish, except for the coordinate time, i.e. 

\begin{equation}
\frac{dx_i}{d\tau} = 0  \qquad \textrm{and}   \qquad    \frac{dx_0}{d\tau} = -e^{\mu/2}.
\end{equation}

So the nonzero components of the energy-momentum tensor of a static perfect fluid with spherical symmetry are

\begin{equation}\label{emtensor-2}
\begin{aligned}
T_{00} &= \rho_0e^{\mu}, \qquad & T_{11} &= p_0e^{\nu}, \\
T_{22} &= p_0r^2, \qquad & T_{33} &= p_0r^2sin^2\theta .
\end{aligned}
\end{equation}

Now, if we compare (\ref{emtensor-1}) and (\ref{emtensor-2}) we will have
\begin{align}
  \rho_0 &= \frac{e^{-\nu}}{8\pi r^2}(r\nu' - 1 + e^{\nu})  \label{density}        \\  
  p_0 &= \frac{e^{-\nu}}{8\pi r^2}(r\mu' + 1 - e^{\nu})  \label{pressure-1}  \\
  p_0 &= \frac{e^{-\nu}}{8\pi r^2}\Big\{\frac{r}{2}(\mu'-\nu') + \frac{r^2}{2}(\mu'' + \frac{\mu'^2}{2} - \frac{\mu'\nu'}{2})\Big\}   \label{pressure-2}     
\end{align}
These are the direct relations between the thermodynamic and metric quantities. For later use, we derive another relation by adding (\ref{density}), (\ref{pressure-1}), and two times (\ref{pressure-2}) 
\begin{equation}
   (\rho_0 + 3p_0) = \frac{e^{-\nu}}{8\pi}\Big(\frac{2\mu'}{r} + \mu'' + \frac{\mu'^2}{2} - \frac{\mu'\nu'}{2} \Big) \nonumber
\end{equation}
which can also be written in the following form 
\begin{equation}\label{usefuleq-1}
   8\pi(\rho_0 + 3p_0)e^{\frac{\mu+\nu}{2}}r^2 = e^{\frac{\mu-\nu}{2}}\Big(\frac{2\mu'}{r} + \mu'' + \frac{\mu'^2}{2} - \frac{\mu'\nu'}{2}\Big)r^2 = \frac{d}{dr}(e^{\frac{\mu-\nu}{2}}\mu'r^2). 
\end{equation}
Also by combining (\ref{density}), (\ref{pressure-1}), and  (\ref{pressure-2})
\begin{equation}\label{usefuleq-2}
    \frac{dp_0}{dr} = -\frac{\rho_0 + p_0}{2}\frac{d\mu}{dr}
\end{equation}
which is the relativistic analogue of the Newtonian relation between pressure and gravitational potential $\varphi$ 
\begin{equation}
    \frac{dp}{dr} = -\rho\frac{d\varphi}{dr}.
\end{equation}

Now we are ready to investigate some important case studies of the Tolman's GRT.

\subsubsection{The Curious Case of Thermal Equilibrium}
The most interesting application of Tolman's GRT is probably on the thermodynamic  equilibrium. Tolman recalls that according to the special theory of relativity any kind of energy has inertia, and hence the equivalence principle implies that it has weight. Then he wonders that, in order to prevent the flow of heat from regions of higher to those of lower gravitational potential, if a temperature gradient is necessary in thermal equilibrium. Then he proves that his intuition is true and if his GRT laws are flawless then there should really be a temperature gradient across a system merged in a gravitational field in equilibrium. As a result he deduces a new and very interesting, yet plausible , concept of \textit{thermodynamic equilibrium}.
\\\\
In his treatment, Tolman first takes a relativistic machanical approach to obtain the condition of thermal equilibrium in a gravitational field, without any resort to thermodynamics. Then he switches to his own theory of GRT, and shows that both approaches lead to the same conclusions. Although the mechanical method is highly instructive, in order to keep ourselves within the scope of this review, we continue on the thermodynamic one.
\\\\
In what follows, first we introduce in details the general case of thermal equilibrium in a gravitational field, according to the GRT. In the next subsection we apply it to the case of the black body radiation, that is the final goal of this article.  
\\\\
 With a glimpse on the definition of entropy vector (\ref{entropy_vector}), it is clear that we can define the entropy of any system by

\begin{equation}\label{entropy}
    S = \int(\phi_0\sqrt{-g}\frac{dx^0}{d\tau})dx^1dx^2dx^3.
\end{equation}
\\
Assuming an \textit{adiabatic} system, there is no flux of matter neither flow of heat through the boundary. So using the second law of GRT (\ref{second_law}) it can easily be shown that 

\begin{equation}
    \frac{\partial}{\partial x^0}\int(\phi_0\sqrt{-g}\frac{dx^0}{d\tau})dx^1dx^2dx^3 \geq 0
\end{equation}
that is equivalent to
\begin{equation}
    \delta S \geq 0.
\end{equation}
\\
Thus, the entropy of an adiabatic system can only increase or remain constant in the timelike coordinate $x^0$. This means that for a static system in thermodynamic equilibrium the entropy is at maximum, i.e. $\delta S = 0$, under the conditions $\delta g_{\mu\nu} = 0$ and $\delta(\partial g_{\mu\nu}/\partial x^\alpha) = 0$. 
Now, consider a \textit{static} system having \textit{spherical symmetry}. We adopt again the line element (\ref{line_element}) and metric components (\ref{metric}). 
\\\\
In case of an adiabatic system

\begin{equation}
    \frac{dx^i}{d\tau} = 0 \qquad   \textrm{and then} \qquad      \frac{dx^0}{d\tau} = \frac{dt}{d\tau} = e^{-\mu/2}.
\end{equation}
\\
Consequently, in thermal equilibrium

\begin{equation}
    \delta\int\phi_0e^{\nu/2}r^2sin\theta drd\theta d\phi = 0 \quad \textrm{with} \quad
\left \{
\begin{tabular}{cc}
$\delta\mu = 0 \quad \delta\mu' = 0$ \\
$\delta\nu = 0 \quad \delta\nu' = 0$
\end{tabular}
\right \}   
\end{equation}

at the boundary, while $'=d/dr$.
For a spherical shell between $r_1$ and $r_2$

\begin{equation}\label{shellequilibrium}
    \delta(4\pi\int_{r_1}^{r_2}\phi_0e^{\nu/2}r^2dr) = 0 \quad \textrm{with} \quad
\left \{
\begin{tabular}{cc}
$\delta\mu = 0 \quad \delta\mu' = 0$ \\
$\delta\nu = 0 \quad \delta\nu' = 0$
\end{tabular}
\right \}  \textrm{at $r_1$ and $r_2$}  
\end{equation}
\\
This means that $\delta S_0 = 0$ with

\begin{equation}\label{shellentropy}
 S_0 = 4\pi\int_{r_1}^{r_2}\phi_0e^{\nu/2}r^2dr.
\end{equation}
\\
If the proper spatial volume is denoted by $V_0$, then the proper four-volume element is $dV_0d\tau = \sqrt{-g}dx^0dx^1dx^2dx^3$. So in the spherical coordinates and having spherical symmetry 

\begin{equation}
    dV_0 = e^{\frac{\mu+\nu}{2}}r^2sin\theta \frac{dt}{d\tau}drd\theta d\phi \quad ,\textrm{with} \quad \frac{dt}{d\tau} = e^{-\mu/2}.
\end{equation}

So for the shell

\begin{equation}
    V_0 = 4\pi\int_{r_1}^{r_2}e^{\nu/2}r^2dr.
\end{equation}

On the other hand, the first law of classical thermodynamics asserts that

\begin{equation}
       \delta S_0 = \frac{\delta U_0}{T_0} + \frac{p_0}{T_0}\delta V_0
\end{equation}
\\
while the subscripts represent the proper value of each quantity, and $U_0 = \rho_0V_0$ is the proper internal energy. Thus in equilibrium

\begin{equation}
    4\pi\int_{r_1}^{r_2}\Big \{\frac{\delta (\rho_{0}e^{\nu/2})}{T_0} + \frac{p_0}{T_0}\delta (e^{\nu/2})\Big \}r^2dr =0 
\end{equation}
\\
Now we substitute for $\rho_0$ from (\ref{density}), perform variations, do integration by parts, and cancel out the terms on the boundaries (since $\delta \nu = 0$ and $\delta\nu' = 0$ at $r_1$ and $r_2$). Then we have

\begin{equation}
    \frac{d}{dr}\Big\{e^{\nu/2}r^2\frac{d}{dr}(\frac{1}{T_0})\Big\} = \frac{4\pi (\rho_{0}+3p_0)}{T_0}e^{\nu/2}r^2
\end{equation}
\\
Using (\ref{usefuleq-1}) we have
\begin{equation}
    \frac{d}{dr}\Big\{e^{\nu/2}r^2\frac{d}{dr}(\frac{1}{T_0})\Big\} = \frac{e^{-\mu/2}}{T_0}\frac{d}{dr}\Big\{e^{-\nu/2}r^2\frac{d}{dr}(e^{\mu/2})\Big\}
\end{equation}
and taking the first integral we obtain 
\begin{equation}
    \frac{d\ln T_0}{dr} = -\frac{1}{2}\frac{d\mu}{dr} + A\frac{e^{-\frac{\mu+\nu}{2}}}{r^2}T_0
\end{equation}
\\
where A is a constant of integration. On basis of physical arguments, we expect that at $r=0$: $dT_0/dr = 0$, $dp_0/dr = 0$, and $T_0 \neq 0$. This implies that $A=0$. Now along with (\ref{usefuleq-2}), to see the general nature of the result we have

\begin{equation}
   \frac{d\ln T_0}{dr} = -\frac{1}{2}\frac{d\mu}{dr} = \frac{1}{\rho_0+p_0}.\frac{dp_0}{dr} 
\end{equation}
\\
That is, decreasing $r$ (going to lower gravitational potentials) results in increasing proper temperature and pressure. 
\\\\
Finally we conclude that

\begin{equation}
    T_0 = Ce^{-\mu/2}
\end{equation}
where C is a constant of integration, or 
\begin{equation}\label{Tolman}
    T_0e^{\mu/2} = T_0\sqrt{-g_{00}} = const.
\end{equation}

This conclusion is very interesting and significant. It states that 
\\
\textit {the proper temperature of a fluid sphere in thermal equilibrium is not constant throughout, but it varies with gravitational potential, increasing with depth toward the center of the sphere.}
\\\\
However, this effect is extremely small for normal macroscopic bodies around us. For example, on the Earth's surface

\begin{equation}
    \frac{d\ln T_0}{dr} \simeq -10^{-18} \quad cm^{-1} 
\end{equation}
\\
While at just one centimeter above the event horizon of a solar-mass black hole

\begin{equation}
    \frac{d\ln T_0}{dr} \simeq -1 \quad cm^{-1} 
\end{equation}
\\
So this variation is important and unignorable only in some astrophysical setups and in gedanken experiments.

\subsubsection{The Black Body Radiation}
As we previously explained, the black body radiation has always been a testbed for various theories of thermodynamics. So it is natural to ask how the black body radiates in gravitational relativistic thermodynamic equilibrium. As the pure black-body radiation is a perfect fluid\footnote{The applicability of the energy-momentum tensor of a perfect fluid to the black-body radiation is justified by the fact that in relativistic mechanics any system whose local properties can be specified by the two scalars $\rho_{0}$ and $p_0$ has such an energy-momentum tensor. And this is surely the case for the black-body radiation. A more rigorous reasoning is based on the electromagnetic nature of the black-body radiation, using the covariant form of the Maxwell's equations \cite{tolman-5}.}, so the same result that we obtained in the previous part is also applicable to it. However, to get some more insight on this case, and also as a pedagogical treatment, we render it directly and independently.
\\\\
Again, although Tolman uses also the formalism of relativistic mechanics for the case of black-body radiation, we concentrate here to his thermodynamic approach.
In the framework of Tolman's GRT we approach this problem in the following manner.
\\\\
According to the classical theory of black-body radiation, for the proper energy density and proper pressure we have

\begin{equation}\label{SB}
    \rho_0 = aT_0^4,   \qquad    p_o = \frac{1}{3} aT_0^4
\end{equation}
where $a$ is the Stefan-Boltzmann constant. So, making use of the first law of classical thermodynamics, for the proper density of entropy we obtain
\begin{equation}
    \phi_0 = \frac{4}{3}aT_0^3
\end{equation}
If we substitute for $T_0$ using (\ref{SB}), then
\begin{equation}
    \phi_0 = \frac{4}{3}a^{1/4}\rho_0^{3/4}
\end{equation}
Now, using (\ref{density}) to connect the density of entropy with the metric we can write
\begin{equation}
    \phi_0 = \frac{4a^{1/4}}{3(8\pi)^{3/4}}e^{-3\nu/4}r^{-3/2}\Big [r\nu'- 1 + e^{\nu}\Big ]^{3/4}
\end{equation}
Using this density of entropy in (\ref{shellequilibrium}) we find the equilibrium condition for a spherical shell in terms of the metric variable $\nu$
\begin{equation}
    \delta\int_{r_1}^{r_2}\Big [r\nu'- 1 + e^{\nu}\Big ]^{3/4}e^{-\nu/4}r^{1/4}dr = 0 
\end{equation}
under the conditions
\begin{equation}
    \delta \nu = 0  \qquad \textrm{and}  \qquad \delta \nu' = 0  \qquad  (\textrm{at}\; r_1 \;\textrm{and} \; r_2).
\end{equation}
Now we take the variations, integrate by parts, and using the boundary conditions cancel out the terms on the limits to come to 
\begin{equation}
    \frac{d}{dr}\Big \{e^{-\nu/2}r^2\frac{d}{dr}\Big(\frac{1}{\rho_{0}^{1/4}}\Big)\Big \} = 8\pi \rho_0^{3/4}e^{\nu/2}r^2
\end{equation}
and using the Stefan-Boltzmann's law (\ref{SB})
\begin{equation}
    \frac{d}{dr}\Big \{e^{-\nu/2}r^2\frac{d}{dr}\Big(\frac{1}{T_0}\Big)\Big \} = 8\pi \frac{\rho_0}{T_0}e^{\nu/2}r^2.
\end{equation}

Noting that for pure radiation  $\rho_0 = 3p_0$, and substituting from (\ref{usefuleq-1})

\begin{equation}
    \frac{d}{dr}\Big \{e^{-\nu/2}r^2\frac{d}{dr}\Big(\frac{1}{T_0}\Big)\Big \} = \frac{e^{-\mu/2}}{T_0}\frac{d}{dr}\Big \{e^{-\nu/2}r^2\frac{d}{dr}(e^{\mu/2})\Big \}
\end{equation}
or
\begin{equation}
    \Big(\frac{1}{T_0}\Big)^{-1}\frac{d}{dr}\Big \{e^{-\nu/2}r^2\frac{d}{dr}\Big(\frac{1}{T_0}\Big)\Big \} = (e^{\mu/2})^{-1}\frac{d}{dr}\Big \{e^{-\nu/2}r^2\frac{d}{dr}(e^{\mu/2})\Big \}.
\end{equation}
A particular solution is
\begin{equation}
    T_0 = Ce^{-\mu/2}
\end{equation}
where $C$ is a constant of integration. This is exactly the same result we obtained for the general case in the previous section.
\\\\
It is worthwhile to mention that in another paper \cite{tolman-5} Tolman and Ehrenfest showed that the above results are also valid in the case of a general static gravitational field containing fluid as well as solid parts.

\section{Some Applications of General Relativistic Thermodynamics}\label{GRT_apps}
One of the theoretical applications of Tolman's GRT may be to determine the local temperature, i.e. the temperature as measured by any observer in her local rest frame, of a radiation field in a general relativistic environment. Generally, it is prohibitively difficult to find solutions for the electromagnetic field in a general relativistic settings, essentially due to the severe nonlinearity of the Einstein's equations (for example see Birrell and Davies, 1982\cite{birrell}). So, people are used to (or compelled to) find asymptotic solutions for such problems. 
This is mainly the playground of the quantum field theory in curved space.
\\\\
The standard method to approach this problem is that the radiation field is considered as a quantum scalar field in a curved space background, because the spin and polarization of photons are irrelevant and effectless in this problem. In this way we get rid of undue calculational complexities. So the wave equation for a scalar field  $\phi(t,\bf{x})$ (the Klein-Gordon equation) which has the form
\begin{equation}\label{KG_SR}
 (\Box + m^2)\phi = 0  
\end{equation}
in the Minkowski flat spacetime (where $\Box = \partial^{\mu}\partial_{\mu} = -\partial^2_t + \partial^2_{\bf{x}}$ is the flat-space second order differential operator and $m$ is the mass of the field quantum) converts to
\begin{equation}\label{KG_GR}
 (\Box + m^2 + \xi R)\phi = 0  
\end{equation}
in the curved spacetime, where $\xi$ is the coupling constant of the field  to the background and $R$ is the Ricci scalar. Here $\Box =\nabla^\mu\nabla_\mu$ is the covariant d’Alembertian operator. We usually assume no interaction with the background, i.e. $\xi=0$, although for the most interesting case of the Schwarzschild spacetime the third term automatically vanishes because then $R = 0$. Also, we consider massless scalar ``photons". So altogether we have
\begin{equation}\label{KG}
\Box\phi = g^{\mu\nu}\nabla_\mu\nabla_\nu\phi = \frac{1}{\sqrt{-g}}\partial_{\mu}(\sqrt{-g}g^{\mu\nu}\partial_{\nu}\phi) = 0.   
\end{equation}
A solution of this equation is the complete orthonormal set of the eigenfunctions of the field (the mode functions) $u_{\mathbf{k}}(t,\mathbf{x}) \propto e^{i(\mathbf{k.x}-\omega t)}$. So the scalar field can be expanded in terms of them as
\begin{equation}\label{expansion}
\phi(t,\mathbf{x}) = \sum_{\mathbf{k}}(a^{}_{\bf{k}}u^{}_{\bf{k}} + a^{\dagger}_{\bf{k}}u^{*}_{\bf{k}})
\end{equation}

with $\omega = \abs{\bf{k}}$ in this case. To quantize the field canonically, one considers the expansion coefficients as operators $\hat{a}$ and $\hat{a}^{\dagger}$ (the annihilation and creation operators, respectively) on the Fock space of the field quantum states , with the canonical commutation relations
\begin{equation}\label{commutations}
[\hat{a}^{}_{\bf{k}},\hat{a}^{}_{\bf{k'}}] = 0, \qquad [\hat{a}^{\dagger}_{\bf{k}},\hat{a}^{\dagger}_{\bf{k'}}] = 0, \qquad  [\hat{a}^{}_{\bf{k}},\hat{a}^{\dagger}_{\bf{k'}}] = \delta_{\bf{kk'}}.
\end{equation}
The vaccum state $\ket{0}$ of the field is defined as the eigenstate for which $\hat{a}\ket{0} = 0$, and the whole Fock space is then built upon it.
On the other hand, an observer needs a ``thermometer'' to measure the temperature of this field. One of the favorite thermometers in this context is the Unruh-DeWitt detector\cite{unruh-1,dewitt}. The Unruh-DeWitt detector is a two-level atom (kind of a qubit, in the jargon of quantum information theory)  having only a monopole interaction with the scalar field, and no angular momentum exchange. Its monopole moment operator is $\hat{\mu}(\tau) = \hat{\sigma}^+e^{i\tau\Delta E} + \hat{\sigma}^-e^{-i\tau\Delta E}$, where $\tau$ is the proper time of the detector, $\Delta E = E_a - E_b$ where $E_a$ and $E_b$ ($E_a > E_b$) are energies of the atomic levels, and $\hat{\sigma}^+$ and $\hat{\sigma}^-$ are raising and lowering ladder operators of the atom, respectively. Suppose that the trajectory of the detector in spacetime is $x^\mu(\tau)$. Then the interaction Hamiltonian of the field-detector system in the detector's rest frame is

\begin{align}\label{hamiltonian}
 \hat{V}(\tau) &= g\hat{\mu}(\tau)\hat{\phi}(x^\mu(\tau))  \nonumber       \\
   &= g\sum_{\bf{k}}(\hat{\sigma}^+e^{i\tau\Delta E} + \hat{\sigma}^-e^{-i\tau\Delta E})(\hat{a}^{}_{\bf{k}}u^{}_{\bf{k}} + \hat{a}^{\dagger}_{\bf{k}}u^{*}_{\bf{k}})
\end{align}

where $g$ is the coupling constant. Provided that $g$ is small, the first-order perturbation theory is used to find the probability of each term of the Hamiltonian (the Fermi's golden rule). For instance, we can find the probability of the process of emission of a photon of some mode by the excited atom into the field vacuum  $(\ket{0,a} \longrightarrow \ket{1_{\bf{k}},b})$ as

\begin{equation}
  P_{em} = \frac{1}{\hbar^2}\abs{\int d\tau\bra{1_{\bf{k}},b}\hat{V}(\tau)\ket{0,a}}^2  
\end{equation}

and job done. Now, if the obtained probability is in the form of Planck distribution, one can easily extract an equilibrium temperature for the atom-field system. However, life is not that easy.  There are two technical bottlenecks in this approach. First, to solve (\ref{KG}) we have to find the positive-frequency Green functions (the Wightman functions). This is a challenging task, even for the simplest cases \cite{louko}. Second, while in the flat Minkowski spacetime a unique vacuum state for all inertial observers is identified, in curved spacetimes the notion of vacuum is ambiguous \cite{fulling}. In general, there is no unique expansion like (\ref{expansion}), and therefore no consensus on the ``no-particle" state between different observers. This, in its turn, makes the notion of ``particle" ambiguous, too. 
These obstacles are traditionally somehow remedied by resorting to some asymptotic or approximate solutions. For example, we may more easily find the Wightman functions and unique vacuum states for inertial frames or far away observers . However, the questions on the behavior of the field (eg. the local temperature) for an arbitrary observer or an arbitrary spacetime trajectory of the detector  remains unanswered.
\\\\
In order to figure out how we can use Tolman's GRT to bypass the above-mentioned difficulties, we deal with the most important and interesting problems of the quantum field theory in curved space: Hawking radiation, and Unruh effect.

\subsection{The Hawking Radiation}

Quite in contrast to the long belief that a black hole can not radiate, Hawking  showed in 1974 that if one regarded the quantum effects near the event horizon, then there would exist a black body radiation of particles that to a ``far away observer" had an entropy proportional to the surface area of the event horizon and (inevitably) a temperature (Hawking temperature), namely\cite{hawking-1,hawking-2}

\begin{equation}\label{Hawking_temperature}
T_H = \frac{\hbar}{2\pi ck_B}\kappa
\end{equation}

where $k_B$ is the Boltzmann constant, and $\kappa = \frac{c^4}{4GM}$ is the surface gravity of the black hole.
Now, we bring forth the question: What will be the Hawking temperature measured by an observer at any other point outside the event horizon? This is a nice instantiation of the black body radiation in a gravitational field which is (according to the equivalence principle) locally equivalent to an accelerating frame. As mentioned above, finding an exact answer to this question is extremely difficult.  However, using the Tolman's relation (\ref{Tolman}) we can bypass the difficulties. Imagine a nonrotating black hole with the Schwarzschild metric

\begin{equation}\label{Schwrzschild_metric}
ds^2 = -(1 - \frac{2M}{r})dt^2 + \frac{dr^2}{1 - \frac{2M}{r}} + r^2d\Omega^2.
\end{equation}

So $g_{00} = -(1 - 2M/r)$. The Hawking temperature at a fixed point at a coordinate radius $r$ is therefore
\begin{equation}\label{local_temp}
T_H(r) = \sqrt{1 - \frac{2M}{r}}T_H.
\end{equation}

This result is exact, with no approximation, and it approaches correctly to the Hawking temperature for distant observers.

\subsection{The Unruh Effect}

Soon after the discovery of the Hawking radiation, probably inspired by the principle of equivalence, some people tried an equivalent (or dual) idea: If a stationary observer near an event horizon of a gravitational field feels herself in a thermal  bath of a black body radiation, then an observer with uniform acceleration in the flat Minkowski spacetime should somehow feel the same. Specifically, while the field is in its vacuum state in the Minkowski space, an accelerated atom senses  a thermal state in the form of a black body radiation. This line of thought started with a paper by Fulling (1973)\cite{fulling}, developed by Davies(1975)\cite{davies}, and culminated by Unruh(1976)\cite{unruh-1}. The equilibrium temperature that is sensed by an observer with uniform proper acceleration $\alpha$ is

\begin{equation}\label{UnruhTemperature}
T_U = \frac{\hbar}{2\pi ck_B}\alpha
\end{equation}

or $T_U = \alpha/2\pi$ in natural units, and is called Unruh temperature. 
Notwithstanding that the Unruh effect has yet not been experimentally verified, it is now generally  accepted. The prevailing interpretation of the Unruh effect is that the thermal bath is not an atomic illusion, but there is a real radiation out there, measurable by other observers including the Minkowski observer\cite{unruh-2}. This is called Unruh or acceleration radiation. However, it is under debate by a few people (eg. Grove, 1986\cite{grove}; Raine, \textit{et al.}, 1991\cite{raine}; Sciama, \textit{et al.}, 1981\cite{sciama}; and Ford, \textit{et al.}, 2006\cite{ford-2}). They believe that although the Unruh effect is a physical reality, the acceleration radiation does not exist and it is just a fictitious effect due to the changes in the internal structure of the detector induced by the accelerating external force.
\\\\
So our question, that is related to the theme of this \textit{Report}, is a conditional one: If the acceleration radiation is true, then what will be the temperature of the thermal bath as measured by inertial Minkowski observers? (For not making the problem more complicated, let's assume the same temperature for all inertial detectors, say we take the Landsberg's theory explained in section \ref{Landsberg}).
\\\\
To answer this question formally, we can transform the mode functions of the thermal bath of the accelerating observer to those of the Minkowski observer (the Bogoliubov transformation) and then obtain the temperature using the same approach as mentioned above. However, we again take the shortcut of the Tolman's relation (\ref{Tolman}). 
\\\\
First, we should distinguish the component $g_{00}$ of the metric of the accelerating frame. As is well known from the theory of special relativity, the transformation from the flat Minkowski Cartesian coordinates $(t,x,y,z)$ to the coordinates $(\eta,\xi,y,z)$ of a frame uniformly accelerated  along the $x$- axis (the Rindler coordinates) is

\begin{equation}\label{Rindler_coo}
\begin{aligned}
t &= \frac{1}{a}e^{a\xi}sinh(a\eta)     \\
x &= \frac{1}{a}e^{a\xi}cosh(a\eta)    
\end{aligned}
\end{equation}
(for $(x>|t|$), where $a>0$ is a constant. The trajectory of an accelerated observer in Rindler coordinates is \cite{carroll}

\begin{equation}\label{trajectory}
\begin{aligned} 
\eta(\tau) &= \frac{\alpha}{a}(\tau)       \\
\xi(\tau) &= \frac{1}{a}ln\Big(\frac{a}{\alpha}\Big)
\end{aligned}
\end{equation}

where $\tau$ is the observer's proper time. We see that for an observer with a uniform acceleration $\alpha$ (the Rindler observer)  $\xi$ is constant and $\alpha = ae^{-a\xi}$. The line element of the flat Minkowski space in the Rindler coordinates is

\begin{equation}\label{Rindler_metric}
ds^2 = e^{2a\xi} (-d\eta^2 + d\xi^2) + dy^2 + dz^2
\end{equation}

So for any uniformly-accelerated observer $g_{00} = -e^{2a\xi}$. Using the Tolman's relation we have

\begin{equation}\label{inertial_temp}
T_U^0 = \sqrt{-g_{00}} T_U = \frac{a}{2\pi} 
\end{equation}

So the temperature of the thermal radiation as sensed by the Minkowski observers (if it exists) is $a/2\pi$. Incidentally, this is the Unruh temperature for the Rindler observer sitting at the origin $\xi = 0$, that is the observer with acceleration $a$. Other Rindler observers with proper acceleration $\alpha$ will measure a ``redshifted'' version of this temperature ($\sqrt{-g_{00}}$ is called the redshift factor). All observers with an acceleration $\alpha > a$ ($\xi < 0$) feel a larger \textit{local} Unruh temperature, and vice versa.

\section{Conclusions}\label{conclusions}
The problem of radiation of a moving black body is delegated to the problem of the absolute temperature of a moving body. 
If we accept that the form of the Planck distribution for a black body is valid in all reference frames, the only parameter that should be replaced by its relativistic counterpart is the temperature. This is, in its turn, a core problem of the theory of relativistic thermodynamics. Although special relativistic thermodynamics has been considered from the very first days of the appearance of the theory of special relativity by experts such as Planck and Einstein, it is astonishingly not quite settled even now and there is no particular formulation of relativistic thermodynamics upon which the experts agreed. 
\\\\
However, focusing on the transformation of temperature between inertial reference frames, existent theories of special relativistic thermodynamics can be bundled in three distinct categories:
\\\\
\hspace*{40pt} 1) $T = T_0/\gamma$ \qquad   (mainly by Planck and Einstein)   \\
\hspace*{40pt} 2) $T = T_0\gamma$  \qquad \;  (mainly by Ott and Arzeli\`{e}s)   \\
\hspace*{40pt} 3) $T = T_0$      \qquad  \;\;\;   (mainly by Landsberg) \\

We may also add the later view of Landsberg that temperature can not be transformed at all and its meaning is just that which is measured by a comoving thermometer. Also a manifestly covariant thermodynamics, presented by van Kampen and Israel, is a promising candidate for a robust theory of special relativistic thermodynamics. However, in its current form it is unfortunately silent about the temperature of uniformly moving bodies.
\\\\
This theoretical turmoil could have been quelled if a specific and discriminating laboratory experiment was performed. But relativistic effects are of the second order and testing and measuring them for macroscopic bodies are formidable tasks that have not technically been achievable as yet.
\\\\
As regards to the general relativistic thermodynamics the theoretical situation seems to be better. The only complete theory is that of Tolman. Tolman's GRT has remarkable and ponderable aspects. In our opinion, the most important one is

\hspace*{20pt}The proper temperature of a body in thermodynamic equilibrium is not \\ 
\hspace*{20pt}constant throughout, but it increases with (the absolute value of) the gravitational potential\\

such that
\begin{equation}
    T_0\sqrt{-g_{00}} = T_M
\end{equation}
where $T_M$ is a constant and obviously the proper temperature at the limit of the Minkowski flat spacetime. For example, for outside a spherically symmetric body with mass $M$
\begin{equation}
    T_0 = T_M\Big(1-\frac{2M}{r}\Big)^{-1/2}.
\end{equation}

On the basis of the Tolman's GRT, the form of the black-body radiation in a gravitational field, which is locally equivalent to an accelerating reference frame, is

\begin{equation}
    \rho(\nu;T) = \frac{8\pi h}{c^3}\frac{\nu^3}{exp(\frac{h\nu}{kT_M/\sqrt{-g_{00}}})-1}.
\end{equation}
\\
Again as an example, in a Schwarzschild spacetime
\begin{equation}
    \rho(\nu;T) = \frac{8\pi h}{c^3}\frac{\nu^3}{exp(\frac{h\nu}{kT_M/\sqrt{(1-2M/r)}})-1}.
\end{equation}
Tolman's GRT can also be of paramount importance in any theory in which, due to computational complexities in finding local temperatures, people recourse to approximate and/or asymptotic solutions. In fact, whenever one has the solutions of the equation of motion of some field for inertial frames or for an observer at infinity, provided that the solutions show an explicit thermal behavior with a specific temperature, the local temperature of any other observer can be directly resulted from the Tolman's relation (\ref{Tolman}). 
Two important examples from quantum field theory were introduced in this \textit{Report}. The Hawking temperature that would be in the form of (\ref{local_temp}) for any stationary observer at any point outside a Schwarzschild black hole, and the temperature of the Unruh radiation (in case of existence) for inertial observers (\ref{inertial_temp}).

\end{document}